\documentclass[aps,prl,twocolumn,superscriptaddress,citeautoscript,floatfix]{revtex4}
\pdfoutput=1
\usepackage{graphicx}
\usepackage{amssymb,amsmath}
\usepackage{subfigure}
\usepackage{epstopdf}

\begin{document}

\title[Draft:\today]{ Quantum motion of charges in $\pi$-conjugated polymers: kinkiness induces localization}

\author{Eric R. Bittner}
\affiliation{Department of Chemistry and Physics, University of Houston, Houston, TX 77204}
\email[Address correspondence to ]{Bittner@UH.edu}

\author{Adam J. Moul\'e}
\affiliation{University of California, Davis,
Department of Chemical Engineering and Materials Science
Davis, CA 95616}

%

\date{\today}

\begin{abstract}
We develop a model for charge trapping on conjugated polymer chains using a continuous 
representation of the polymer.  By constraining the motion to along the chain, we find that kinks along the 
chain serve as points of attraction and can induce localization and spontaneous 
buckling of the chain.   We implement this in a model system of a conjugated polymer donor and fullerene
acceptor to mimic a polymer based photovoltaic system.
We find that geometric fluctuations of the 
polymer give rise to deep potential energy traps at kinks which are often stronger than the Coulomb 
interaction between the electron and hole. 
\end{abstract}

\maketitle

One of the persistent questions arising in the modeling of polymer-based solar cells is how the morphology and dynamics of the polymeric molecules comprising the cell relates to overall efficiency of the cell to separate photo-excitations into mobile charges.  
On the short time-scale, we can consider the charge-separation event in a framework of a fixed nuclear geometry.  For certain, the efficiency and rate of exciton fission is intimately dependent upon how donor and acceptor species are in contact with each other. 
Once exciton fission has occurred to produce a bound charge-transfer pair,
 one needs to consider how the fluctuations and dynamics of polymer chains themselves 
play into further separating this state into mobile charge carriers that can be collected.   However, in organic conjugated polymer semiconductors, the idea of a ``mobile charge'' is somewhat of a misnomer since disorder, strong-electron phonon coupling, and
chemical defects can lead to charge trapping and drastically diminish mobility.\cite{muccini:6296,Ladik:2003}

The underlying questions are  whether or not  the $\pi$-conjugation persistence length 
 stretches over the entire length of a polymer chain, and will the mobility of a free charge on the chain be
  limited by geometry of the 
chain itself.   To address these,  we  consider a conjugated polymer to be a continuous filament in $R^{3}$ 
described by the vector $\vec\gamma(s)$ where $s$ is the 
arc-length on which a mobile charge with mass $m$ is bound via a confining potential. 
While the polymer may be bent and kinked, we assume that the $\pi$ electronic 
system is continuous along the chain and that the particle can freely move from one end to the other.  In the limit that the 
polymer is fully extended, the free-charge would behave as a particle in a one-dimensional box.  


However, when introducing bends and kinks in the polymer, one needs to consider the effect 
of constraining the motion to follow the three-dimensional geometry of the polymer itself.   
In general, the  quantum motion of the charge satisfies the 
time-dependent Schr\"odinger equation $i\hbar \dot\psi(q) = H\psi(q)$ with $H$ given by the kinetic energy operator in 3-dimensions. 
Da Costa showed that these conditions lead to a reduced  Schr\"odinger equation for the longitudinal
motion of a quantum particle 
along $\vec\gamma(s)$.  \cite{PhysRevA.23.1982,PhysRevA.25.2893}  We briefly summarize this result here. 
To constrain the motion to along a filament, 
we define a coordinate frame in terms of the tangent to the curve, $\hat t(s)= \vec\gamma'(s)$, and vectors $\hat n_{2}$ and $\hat n_{3}$. 
These unit vectors define a local orthonormal coordinate frame centered at $\gamma(s)$ and 
are related by rotation, ${\bf T}(\theta)$ about $\hat t$ to the normal and binormal vectors $\hat n$ and $\hat b$. 
Any point in the $\hat n_{2},\hat n_{3}$ plane can be written as 
 $\vec R(s) = \vec\gamma(s) + q_{2}\hat n_{2} + q_{3}\hat n_{3}$.
Vectors $\hat t$, $\hat n$, and $\hat b$ satisfy the Frenet-Serret equations along $\vec \gamma(s)$:
$\vec\gamma'(s) = {\bf  F}\cdot\gamma(s)$ where ${\bf F}$ is the anti-symmetric Frenet-Serret matrix with depends upon the curvature, $\kappa$ and torsion $\tau$.  

Note that the derivative of the 
rotation angle with respect to the arc-length is equal to the torsion, $\theta'  = \tau$. 
Using  $\vec R(s)$, one then finds the covariant metric tensor is diagonal with 
elements  $G_{11} = (1-\kappa f)^{2}$, and $G_{22} = G_{33} = 1$
 where $f = q_{2}\cos\theta(s) + q_{3}\sin\theta(s)$ and $\theta(s)$ is the rotation of the frame about $\hat t$. 
 Including this rotation of the frame relative to the Frenet-Serret frame brings the covariant metric into 
 a diagonal form at all points along the curve.  
Inverting $G_{\mu\nu}$ and writing the kinetic energy operator in terms of the  Laplace-Beltrami operator, 
allows one to construct 
an effective one-dimensional 
Schr\"odinger equation  for the free, but constrained, motion along the curve
\begin{eqnarray}
-\frac{\hbar^{2}}{2m}\left(\frac{\partial^{2}}{\partial s^{2}}  + \frac{\kappa(s)^{2}}{4} \right) \psi(s) =i\hbar \dot\psi(s).\label{curvedse}
\end{eqnarray}

The original goal of da Costa's work was to understand the operator ordering problem when solving 
quantum problems in curved metrics; however,
for the purposes of understanding charge localization and motion in 
conjugated polymer-based devices, it provides a useful way to connect the influence of the geometry of the 
the chain with the dynamics of a mobile charge along the chain.

\begin{figure}[t]
\subfigure[]{\includegraphics[width=0.45\columnwidth]{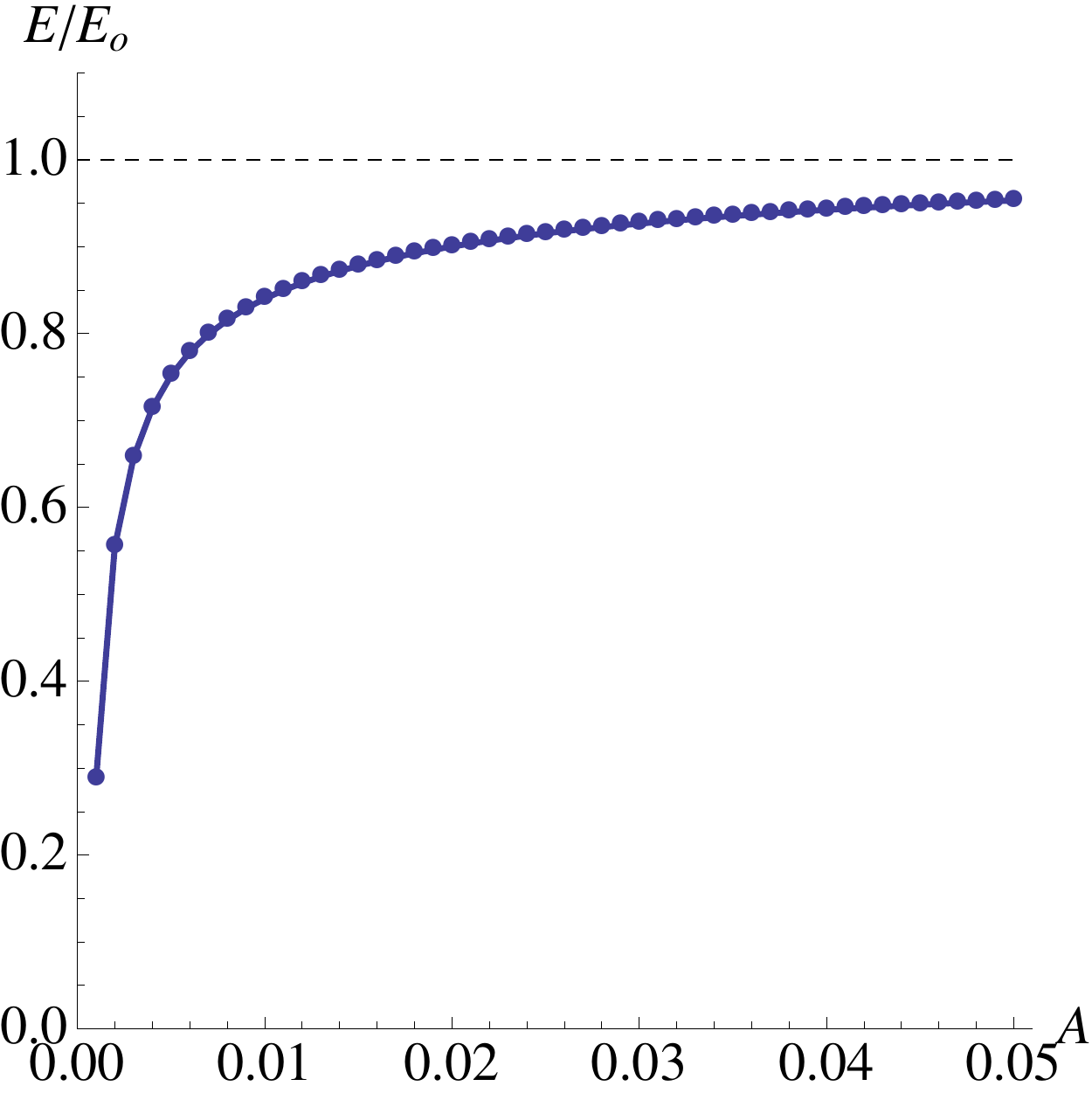}}
\subfigure[]{\includegraphics[width=0.45\columnwidth]{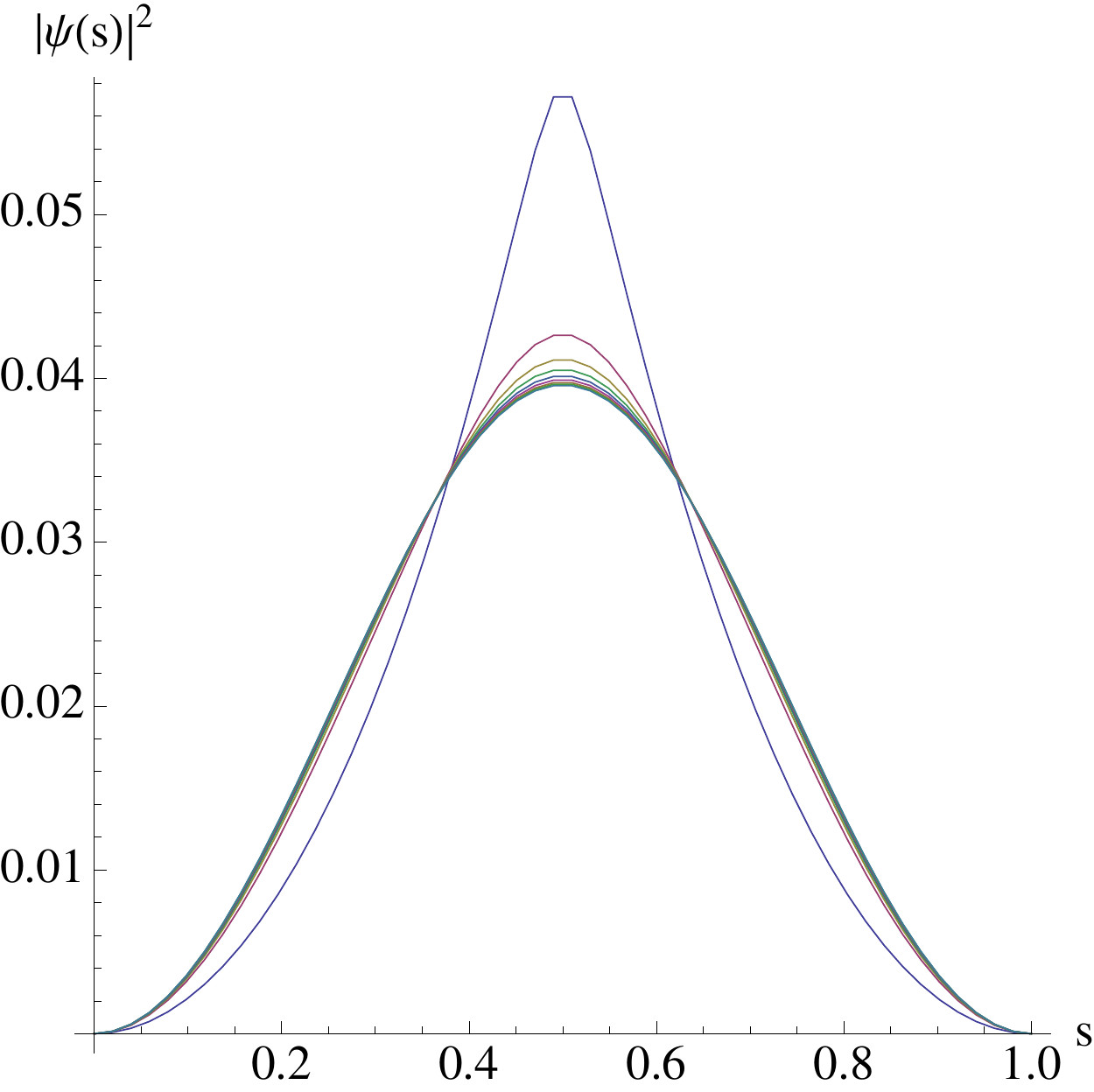}}
\caption{Ground-state energy (a) and probability density (b) for a quantum particle on a flexible parabolic rod with bending modulus $A$.
 Our units are such that $E_{o} =\hbar^{2}\pi^{2}/2mL^{2} = 1$ is the ground state energy of a particle in a 1D box. 
 }\label{fig1}
\end{figure}

Consider the case of a single quantum particle bound to a flexible rod given by, $\vec\gamma(s)$ with 
bending modulus, $A$.  We can write the total ground-state energy is a functional of $\kappa(s)$ and the longitudinal wave function, $\psi(s)$
\begin{eqnarray}
E[\vec\gamma,\psi] = \frac{A}{2}\int_{0}^{L}\kappa^{2}(s)ds + \langle \psi | H[\gamma] | \psi \rangle,\label{estrain}
\end{eqnarray}
subject to the constraint that the length of the rod is fixed  $L = \int_{0}^{L}|\gamma'(s)|ds$  and $\psi$ is normalized to unity with boundary conditions $\psi(0) = \psi(L) = 0$. 
Bending the rod about the center increases the curvature and lowers energy of the quantum particle bound to the rod but increases the strain energy.  At some point, the two 
come into balance and we can minimize the total energy self-consistently.   
As a trial curve, we take a parabolic curve $\vec\gamma(s) = \{a (s-1/2), b(s-1/2)^{2},0\} $ with $a$ and $b$ chosen so that $\gamma(s)$
is a fixed length and minimize the total energy with respect to $\psi$ and the curvature.  
In Fig.~\ref{fig1}a we show energy vs. bending modulus for such a system. The 
dashed line corresponds to the 1D box limit where the rod is perfectly straight.
As the bending modulus decreases, the rod is increasingly bent about its middle and the quantum state becomes increasingly localized as seen in Fig.~\ref{fig1}b.  

This flexing and localization is spontaneous for all values of the bending modulus and should be a universal property of all quasi-one-dimensional quantum systems.  That stated,  $\pi$-conjugated molecular systems and carbon nanotubes are very stiff;  however, it may be possible to see a spontaneous buckling effect in sinusoidal or $S$-shaped systems.  For purposes of this paper, we 
are focused upon how geometric fluctuations in a conjugated polymer chain can contribute to quantum
localization and charge separation in optical-electronic and photovoltaic materials composed of
conjugated polymer molecules.

\begin{figure}[b]
\subfigure[$t=0$]{\includegraphics[width=0.49\columnwidth]{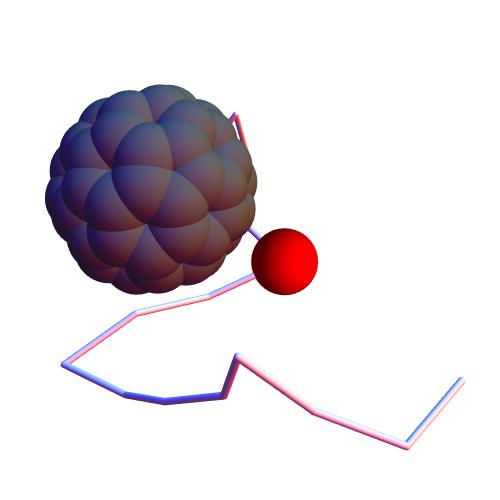}}
\subfigure[$t=100fs$]{\includegraphics[width=0.49\columnwidth]{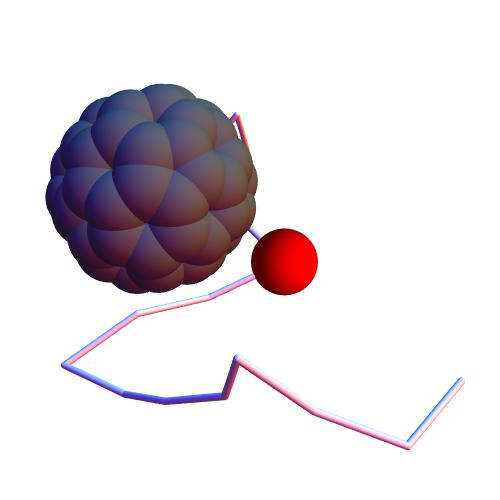}}\\
\subfigure[$t=200fs$]{\includegraphics[width=0.49\columnwidth]{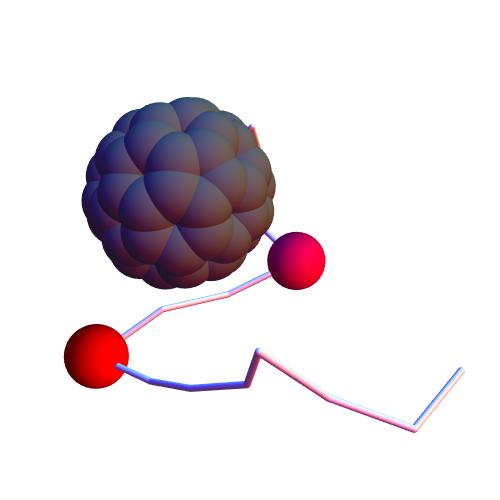}}
\subfigure[$t=300fs$]{\includegraphics[width=0.49\columnwidth]{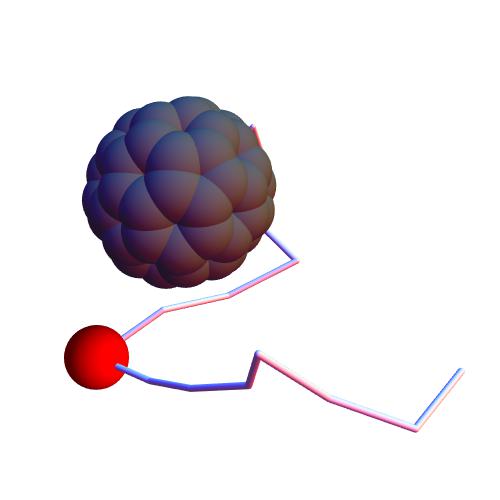}}
\caption{Charge tunneling between kinks on a polymer chain. 
At $t=0$ the  chain takes on a (+) charge and the $C_{60}$ takes on a (-1) charge. The color and magnitude of the sphere on the chain are indicative
of the charge at a given site.  At $t=20fs$,  the quantum state is clearly delocalized between kinks.  At $t=30fs$ the quantum state is localized at a new kink.  }\label{fig2}
\end{figure}

We developed a straightforward model for simulating the dynamics in a 
model bulk heterojunction photovoltaic 
system consisting of $C_{60}$ derivatives blended with poly-thiophene chains.
Such systems and their derivatives are of considerable interest for fabricating highly-efficient photovoltaic devices based upon organic 
materials. 
  \cite{halls:3120,harigaya:13676,Imahori:2004,sariciftci:585,xue:3013}
Since we are primarily interested in the coarse-grained as opposed to atomistic 
motion of the system, we treated the $C_{60}$  as a 
Lennard-Jones (LJ) sphere and the polythiophene chain as linked chain of Lennard-Jones beads with a bending potential between adjacent sites.  
The initial conditions were generated
from a classical molecular dynamics (MD) simulation of a single $C_{60}$ sphere embedded in cell with 25  polymer chains each with 20 sites at constant $T = 300K$ and volume as imposed by 
periodic boundary conditions.  
Upon photo-excitation, we assume that charge transfer occurs in the framework of a 
fixed geometry of the system producing a single  $C_{60}^{-}$ anion and  a single positively charged  chain located in close proximity to the the $C_{60}^{-}$ anion. 
Since the chains are treated as a discrete set of beads, we use a bicubic spline to compute an  interpolation function for $\gamma(s)$ along the chain which is used
for computing the curvature and Frenet-Serret frame along the polymer curve. 
We then assign local site charges by solving the Schr\"odinger equation for a particle 
on a curve (Eq. ~\ref{curvedse}), adding to this a 
site energy determined by the Coulomb interaction between the sites on the chain and the anion, 
\begin{eqnarray}
-\frac{\hbar^{2}}{2m}\psi''   + (V(|\vec R-\vec\gamma(s)|)- \frac{\hbar^{2}}{8m}\kappa^{2})\psi  = E\psi \label{h-reduced}
\end{eqnarray}
where $V(r)$ is the Coulomb interaction between the $C_{60}^{-}$ at location $\vec R$ and an arbitrary location along the chain $\vec\gamma(s)$. 
We solved Eq. 4 numerically using a $n$-point discrete variable representation based upon Tchebychev polynomials.\cite{light:1400,Light:1992} 
Site charges are given by $|\psi(s_{i})|^{2}$ evaluated at the center of each bead.
We assume that the adiabatic approximation remains valid so that the interaction between the quantum and classical degrees of
freedom can be described within the Hellmann-Feynman theorem. 
The information from the quantum calculation was then passed to a molecular dynamics (MD) code (Tinker) \cite{ponder:2004}
and for a single 1 fs MD step.  This cycle was repeated for each dynamical step.  
Numerically, all this was accomplished using Mathematica (v.8.0) \cite{MMa8.0} and the Tinker (v.4.2) molecular dynamics code.   

\begin{figure}[t]
{\includegraphics[width=\columnwidth]{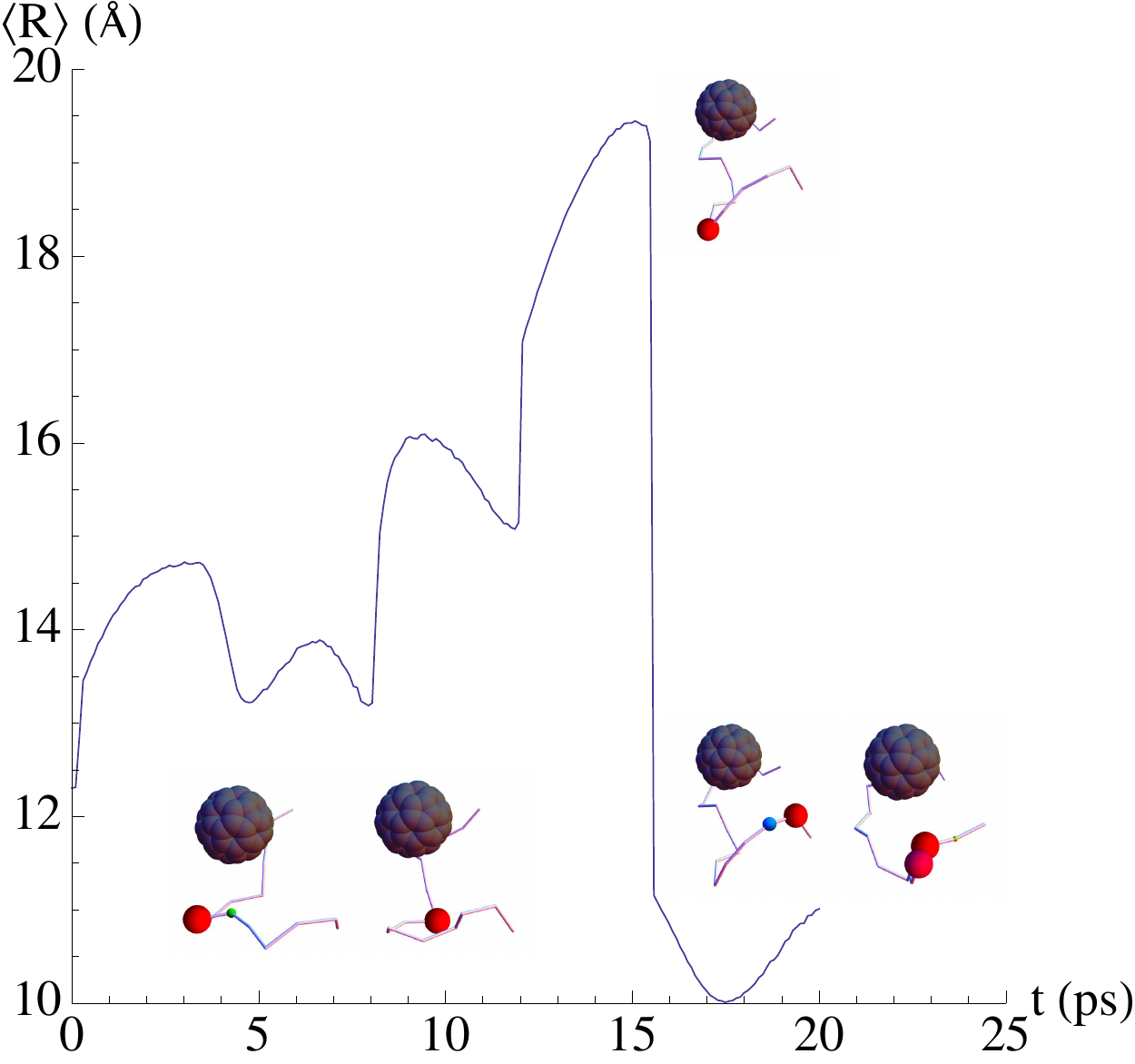}}
\caption{  Distance between the $C_{60}^{-}$ and average + charge location vs. time.  Inset figures show the 
polymer conformation and charge location at selected times.   }\label{fig3}
\end{figure}

\begin{figure*}
\subfigure[]{\includegraphics[width=0.9\columnwidth]{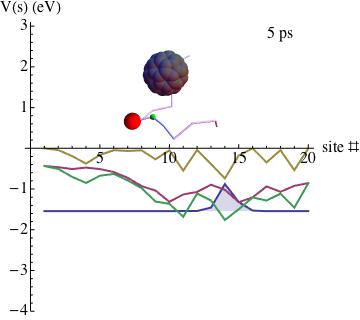}}
\subfigure[]{\includegraphics[width=0.9\columnwidth]{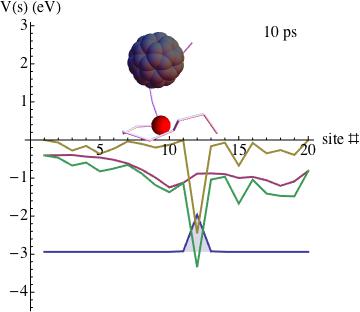}}
\subfigure[]{\includegraphics[width=0.9\columnwidth]{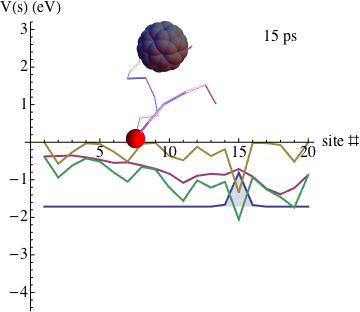}}
\subfigure[]{\includegraphics[width=0.9\columnwidth]{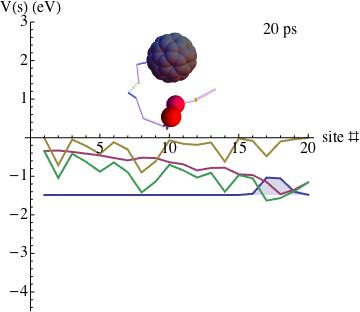}}
\caption{Potential energy contributions and ground-state wave functions at various points along the simulation shown in Fig. 3.
Yellow = curvature potential, Red =Coulomb potential, Green = Total potential, Blue = wave function (scaled and off-set to ground-state energy).}
\label{fig4}
\end{figure*}

In Fig.~\ref{fig2}(a-d) we show a sequence of frames from a representative simulation.  At $t=0$, shown in Fig.~\ref{fig2}a, 
we initiated the quantum dynamics by placing a non-mobile negative charge on the $C_{60}$ sphere and a mobile positive charge
in the lowest energy eigenstate of a nearby polymer chain.  The initial state is largely localized at a kink close to the $C_{60}$.    However, over the 
next 300 fs the charge tunnels from one kink to a nearby kink and becomes localized at the new kink.  
Similar transitions occur repeatedly over the course of a given simulation. 

 In Fig.~\ref{fig3} we show the distance between the positive charge on the chain and the 
$C_{60}^{-}$ over the course of the simulation.  Inset figures give an indication of the geometry  of the chain and the distribution of charges along the chain at selected 
intervals.  In particular, the dramatic jump at 15ps corresponds to the tunneling of the state along the chain towards sites closer to the $C_{60}^{-}$.  

In Fig.~\ref{fig4} we compare the effect of the various potential energy contributions in Eq.~\ref{h-reduced} at the same points of time highlighted in Fig.~\ref{fig3}.  
The ground-state energy and wave function of each are also indicated in these figures.  What is striking is that the curvature contribution (in yellow) is 
generally much less than the Coulomb energy between sites along the polymer chain and the $C_{60}^{-1}$.  However, the presence of 
kinks on the chain give rise to deep traps along the chain that in many cases are stronger than the Coulomb interaction between the charge and the 
$C_{60}^{-1}$ anion.  In fact, in many cases, these kinks move the quantum state away from $C_{60}^{-1}$ anion for long periods of time.  
This implies that in more realistic systems, {\em dynamical fluctuations of the chains themselves may play a key role in the charge separation dynamics of bulk-heterojunction systems. }

{\bf Summary: }
The  geometry of curves gives rise to an effective one-dimensional Schr\"dinger equation for a particle bound to the curve in which the particle is attracted to regions of higher curvature.   This purely a property of the geometry of curves and universal to all quasi-one dimensional quantum systems.   We showed that
when combined with a bending modulus, a particle bound to a linear curve will induce a spontaneous 
symmetry breaking to bend the curve to optimize the curvature. 

We then applied this to develop a coarse-grained model of the charge-localization dynamics 
in a model of a $C_{60}$:thiophene based  photovoltaic material. Combining molecular dynamics of 
the polymers with quantum evaluation of a mobile charge bound to a single polymer we show that kinks within the chain lead to localization and tunneling of the charge along the chain.  We conclude that
kinks and fluctuations of the chain may play an important role in the charge fission and separation 
dynamics in polymer-based photovoltaic materials. 

These model simulations underscore the extent to which the dynamics and morphology of the polymer chain influences charge separation in  polymer-fullerene based cells.  While the current model does not account for quantum chemical effects, such as conjugation breaks due to orbital mis-alignment, chemical defects on the chain, or charge transfer between neighboring chains, 
such effects can be incorporated into the model in a systematic way and will serve the basis for subsequent investigations.

\begin{acknowledgments}
This work was supported by the
National Science Foundation (CHE-1011894) and  Robert A. Welch Foundation (E-1334). 
We also acknowledge conversations with Prof. Delmar Larsen and Roland Faller. 
\end{acknowledgments}


\begin{thebibliography}{13}
\expandafter\ifx\csname natexlab\endcsname\relax\def\natexlab#1{#1}\fi
\expandafter\ifx\csname bibnamefont\endcsname\relax
  \def\bibnamefont#1{#1}\fi
\expandafter\ifx\csname bibfnamefont\endcsname\relax
  \def\bibfnamefont#1{#1}\fi
\expandafter\ifx\csname citenamefont\endcsname\relax
  \def\citenamefont#1{#1}\fi
\expandafter\ifx\csname url\endcsname\relax
  \def\url#1{\texttt{#1}}\fi
\expandafter\ifx\csname urlprefix\endcsname\relax\def\urlprefix{URL }\fi
\providecommand{\bibinfo}[2]{#2}
\providecommand{\eprint}[2][]{\url{#2}}

\bibitem[{\citenamefont{Muccini et~al.}(2000)\citenamefont{Muccini, Schneider,
  Taliani, Sokolowski, Umbach, Beljonne, Cornil, and Bredas}}]{muccini:6296}
\bibinfo{author}{\bibfnamefont{M.}~\bibnamefont{Muccini}},
  \bibinfo{author}{\bibfnamefont{M.}~\bibnamefont{Schneider}},
  \bibinfo{author}{\bibfnamefont{C.}~\bibnamefont{Taliani}},
  \bibinfo{author}{\bibfnamefont{M.}~\bibnamefont{Sokolowski}},
  \bibinfo{author}{\bibfnamefont{E.}~\bibnamefont{Umbach}},
  \bibinfo{author}{\bibfnamefont{D.}~\bibnamefont{Beljonne}},
  \bibinfo{author}{\bibfnamefont{J.}~\bibnamefont{Cornil}}, \bibnamefont{and}
  \bibinfo{author}{\bibfnamefont{J.~L.} \bibnamefont{Bredas}},
  \bibinfo{journal}{Physical Review B (Condensed Matter and Materials Physics)}
  \textbf{\bibinfo{volume}{62}}, \bibinfo{pages}{6296} (\bibinfo{year}{2000}),
  \urlprefix\url{http://link.aps.org/abstract/PRB/v62/p6296}.

\bibitem[{\citenamefont{Ladik}(2003)}]{Ladik:2003}
\bibinfo{author}{\bibfnamefont{J.~J.} \bibnamefont{Ladik}},
  \bibinfo{journal}{Fundamental World of Quantum Chemistry}
  \textbf{\bibinfo{volume}{2}}, \bibinfo{pages}{271} (\bibinfo{year}{2003}).

\bibitem[{\citenamefont{da~Costa}(1981)}]{PhysRevA.23.1982}
\bibinfo{author}{\bibfnamefont{R.~C.~T.} \bibnamefont{da~Costa}},
  \bibinfo{journal}{Phys. Rev. A} \textbf{\bibinfo{volume}{23}},
  \bibinfo{pages}{1982} (\bibinfo{year}{1981}),
  \urlprefix\url{http://link.aps.org/doi/10.1103/PhysRevA.23.1982}.

\bibitem[{\citenamefont{da~Costa}(1982)}]{PhysRevA.25.2893}
\bibinfo{author}{\bibfnamefont{R.~C.~T.} \bibnamefont{da~Costa}},
  \bibinfo{journal}{Phys. Rev. A} \textbf{\bibinfo{volume}{25}},
  \bibinfo{pages}{2893} (\bibinfo{year}{1982}),
  \urlprefix\url{http://link.aps.org/doi/10.1103/PhysRevA.25.2893}.

\bibitem[{\citenamefont{Halls et~al.}(1996)\citenamefont{Halls, Pichler,
  Friend, Moratti, and Holmes}}]{halls:3120}
\bibinfo{author}{\bibfnamefont{J.~J.~M.} \bibnamefont{Halls}},
  \bibinfo{author}{\bibfnamefont{K.}~\bibnamefont{Pichler}},
  \bibinfo{author}{\bibfnamefont{R.~H.} \bibnamefont{Friend}},
  \bibinfo{author}{\bibfnamefont{S.~C.} \bibnamefont{Moratti}},
  \bibnamefont{and} \bibinfo{author}{\bibfnamefont{A.~B.}
  \bibnamefont{Holmes}}, \bibinfo{journal}{Applied Physics Letters}
  \textbf{\bibinfo{volume}{68}}, \bibinfo{pages}{3120} (\bibinfo{year}{1996}),
  \urlprefix\url{http://link.aip.org/link/?APL/68/3120/1}.

\bibitem[{\citenamefont{Harigaya}(1992)}]{harigaya:13676}
\bibinfo{author}{\bibfnamefont{K.}~\bibnamefont{Harigaya}},
  \bibinfo{journal}{Physical Review B (Condensed Matter)}
  \textbf{\bibinfo{volume}{45}}, \bibinfo{pages}{13676} (\bibinfo{year}{1992}),
  \urlprefix\url{http://link.aps.org/abstract/PRB/v45/p13676}.

\bibitem[{\citenamefont{Imahori et~al.}(2004)\citenamefont{Imahori, Sekiguchi,
  Kashiwagi, Sato, Araki, Ito, Yamada, and Fukuzumi}}]{Imahori:2004}
\bibinfo{author}{\bibfnamefont{H.}~\bibnamefont{Imahori}},
  \bibinfo{author}{\bibfnamefont{Y.}~\bibnamefont{Sekiguchi}},
  \bibinfo{author}{\bibfnamefont{Y.}~\bibnamefont{Kashiwagi}},
  \bibinfo{author}{\bibfnamefont{T.}~\bibnamefont{Sato}},
  \bibinfo{author}{\bibfnamefont{Y.}~\bibnamefont{Araki}},
  \bibinfo{author}{\bibfnamefont{O.}~\bibnamefont{Ito}},
  \bibinfo{author}{\bibfnamefont{H.}~\bibnamefont{Yamada}}, \bibnamefont{and}
  \bibinfo{author}{\bibfnamefont{S.}~\bibnamefont{Fukuzumi}},
  \bibinfo{journal}{Chemistry} \textbf{\bibinfo{volume}{10}},
  \bibinfo{pages}{3184} (\bibinfo{year}{2004}), ISSN \bibinfo{issn}{0947-6539
  (Print)}.

\bibitem[{\citenamefont{Sariciftci et~al.}(1993)\citenamefont{Sariciftci,
  Braun, Zhang, Srdanov, Heeger, Stucky, and Wudl}}]{sariciftci:585}
\bibinfo{author}{\bibfnamefont{N.~S.} \bibnamefont{Sariciftci}},
  \bibinfo{author}{\bibfnamefont{D.}~\bibnamefont{Braun}},
  \bibinfo{author}{\bibfnamefont{C.}~\bibnamefont{Zhang}},
  \bibinfo{author}{\bibfnamefont{V.~I.} \bibnamefont{Srdanov}},
  \bibinfo{author}{\bibfnamefont{A.~J.} \bibnamefont{Heeger}},
  \bibinfo{author}{\bibfnamefont{G.}~\bibnamefont{Stucky}}, \bibnamefont{and}
  \bibinfo{author}{\bibfnamefont{F.}~\bibnamefont{Wudl}},
  \bibinfo{journal}{Applied Physics Letters} \textbf{\bibinfo{volume}{62}},
  \bibinfo{pages}{585} (\bibinfo{year}{1993}),
  \urlprefix\url{http://link.aip.org/link/?APL/62/585/1}.

\bibitem[{\citenamefont{Xue et~al.}(2004)\citenamefont{Xue, Uchida, Rand, and
  Forrest}}]{xue:3013}
\bibinfo{author}{\bibfnamefont{J.}~\bibnamefont{Xue}},
  \bibinfo{author}{\bibfnamefont{S.}~\bibnamefont{Uchida}},
  \bibinfo{author}{\bibfnamefont{B.~P.} \bibnamefont{Rand}}, \bibnamefont{and}
  \bibinfo{author}{\bibfnamefont{S.~R.} \bibnamefont{Forrest}},
  \bibinfo{journal}{Applied Physics Letters} \textbf{\bibinfo{volume}{84}},
  \bibinfo{pages}{3013} (\bibinfo{year}{2004}),
  \urlprefix\url{http://link.aip.org/link/?APL/84/3013/1}.

\bibitem[{\citenamefont{Light et~al.}(1985)\citenamefont{Light, Hamilton, and
  Lill}}]{light:1400}
\bibinfo{author}{\bibfnamefont{J.~C.} \bibnamefont{Light}},
  \bibinfo{author}{\bibfnamefont{I.~P.} \bibnamefont{Hamilton}},
  \bibnamefont{and} \bibinfo{author}{\bibfnamefont{J.~V.} \bibnamefont{Lill}},
  \bibinfo{journal}{The Journal of Chemical Physics}
  \textbf{\bibinfo{volume}{82}}, \bibinfo{pages}{1400} (\bibinfo{year}{1985}),
  \urlprefix\url{http://link.aip.org/link/?JCP/82/1400/1}.

\bibitem[{\citenamefont{Light}(1992)}]{Light:1992}
\bibinfo{author}{\bibfnamefont{J.~C.} \bibnamefont{Light}}, in
  \emph{\bibinfo{booktitle}{Time-Dependent Quantum Molecular Dynamics}}
  (\bibinfo{publisher}{Plenum-Press}, \bibinfo{year}{1992}).

\bibitem[{\citenamefont{Ponder}(2004)}]{ponder:2004}
\bibinfo{author}{\bibfnamefont{J.~W.} \bibnamefont{Ponder}},
  \emph{\bibinfo{title}{{TINKER}: {Software} Tools for Molecular Design, 4.2
  ed}} (\bibinfo{year}{2004}), \urlprefix\url{http://dasher.wustl.edu/tinker/}.

\bibitem[{MMa(2010)}]{MMa8.0}
\emph{\bibinfo{title}{Mathematica}} (\bibinfo{publisher}{{Wolfram Research,
  Inc.}}, \bibinfo{address}{Champaign, IL}, \bibinfo{year}{2010}),
  \bibinfo{note}{version 8}.

\end{thebibliography}
\end{document}